# Fluorescence Suppressor Gas Electron Multiplier (FS-GEM)

Fabio Sauli

*CERN, Geneva, Switzerland*

A method is proposed to dump the spurious signals induced by fluorescence of the electrodes in gaseous detectors under strong X-Ray irradiation, introducing a special fluorescence suppressor electrode between the sensitive and the amplifying parts of the detector.

The gas electron multiplier (GEM)(Sauli, 1997), originally developed for the needs of particle physics experimentation, is used in many other applied fields, taking advantage of its performances: sub-mm localization, high acquisition rates, easy of construction and long-term reliability (Sauli, 2016). A problem has appeared however in the detection of high-rate soft X-rays, between a few and a few tens of keV, due to the fluorescence of the metals used for the electrodes, including drift and read-out, breeding secondary photoelectrons in the same range of energies and adding to the signal; copper and chromium, used for manufacturing standard GEMs, are particularly effectual. An analysis of the processes encountered in high-rate fusion plasma imaging is provided for example in Ref. (Chernyshova *et al.*, 2019).

Use of light materials as aluminum and diamond-like carbon (DLC) layers as electrodes has been envisaged to moderate the fluorescence problem; with a K-shell energy below 2 keV, Al and C fluoresce can be easily discriminated. Manufacturing aluminum GEM electrodes has however encountered difficulties with the quality of the artwork, affecting performances; for DLC, one additional constraint is the very high resistivity of the layers restraining high-rate operations.

The present work describes a possible alternative: insertion between the sensitive and the amplifying sections of the detector of a special electrode, named "fluorescence suppressor GEM" (FS-GEM), absorbing the photons generated by the incoming radiation in a multi-GEM structure (Figure 1). Manufactured on a thick glass or fiberglass plate, the FS-GEM has the electrode facing the drift made of a light material, Al or DLC; the other side can be a conventional Cu or Cu-Cr layer.

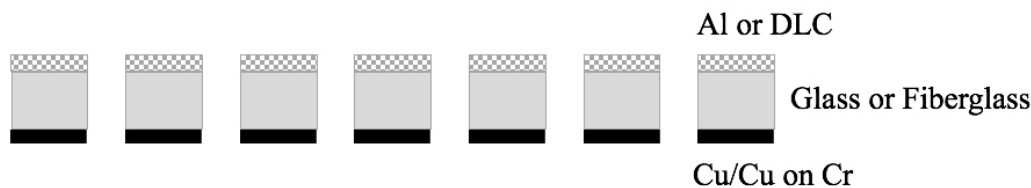

(Figure 1 (not to scale): The proposed special fluorescence suppressor electrode (FS-GEM)

To attain the gain needed to detect the X-ray conversions, the FS-GEM is followed by one or more standard Glass or thin-polymer GEMs; inserted between the sensitive volume and the amplifying section of the detector, the FS-GEM transmits the electrons generated by ionizing events in the drift gap while absorbing fluorescence photons produced by the other electrodes in the stack (Figure 2).

Operating the FS-GEM at a moderate gain serves the double purpose of providing full amplification only to photon conversions in the sensitive volume, while absorbing the fluorescence of other electrodes, and to limit the signal current in the upper electrode to a few percent of the total (Bachmann *et al.*, 1999) thus avoiding problems connected with the use of a high-resistivity DLC layer. As shown in Figure 3, with a 500 µm thick glass, fluorescence photons up to $\sim$ 10-15 KeV are effectively absorbed.

The best option to realize the FS-GEM would to adapt the technology used to manufacture fine-pitch glass GEMs described for example Ref. (Takahashi *et al.*, 2013); this requires however to develop a method to coat one or both sides of the glass with a DLC layer. The holes can then be made by laser drilling, if a wet etching process is not found. Alternatively, the FS-GEM could be manufactured on a fiberglass sheet by mechanical or laser drilling, as done for the Thick-GEM detectors (Chechik *et al.*, 2004). The electrode could be realized directly coating a fiberglass sheet as described in Ref. (Song *et al.*, 2020), or pasting a thin DLC-coated polymer foil as in Ref. (Lv *et al.*, 2020) to a single-sided printed circuit sheet before drilling.

The photon-absorbing electrode can be added to existing systems without affecting their detection efficiency and performances.

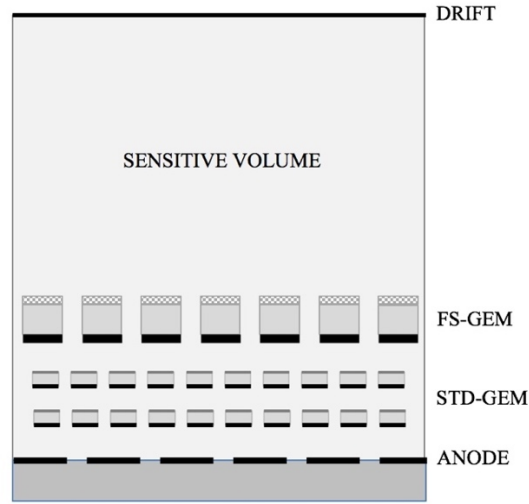

(Figure 2 (not to scale): The fluorescence suppressor GEM inserted between the sensitive volume and a Double-GEM amplifier.

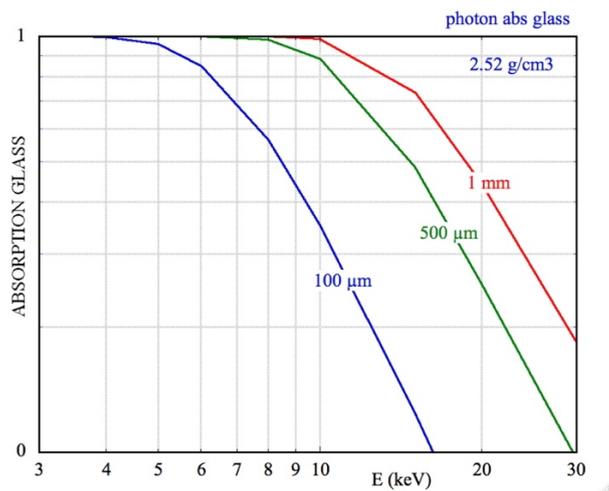

Figure 3: Photon absorption efficiency of glass as a function of energy.